\documentclass[twocolumn,showpacs,prl]{revtex4}

\ifx\pdfoutput\undefined
\usepackage{graphicx}
\else
\usepackage{graphicx}
\fi

\usepackage[center]{subfigure}
\usepackage{bm}
\usepackage{amsmath}
\usepackage{latexsym}

\def\beq{\begin{equation}}
\def\eeq{\end{equation}}
\def\nn{\nonumber}

\begin{document}

\title{Predator-Prey Quasi-cycles from a Path Integral Formalism}
\author{Thomas Butler and David Reynolds}
\affiliation{Department of Physics and Institute for Genomic Biology,
University of Illinois at Urbana Champaign, 1110 West Green Street, Urbana, IL 61801 USA}

\date{\today}

\begin{abstract}

The existence of beyond mean field quasi-cycle oscillations in a simple spatial model of predator prey interactions is derived from a path integral formalism.  The results agree substantially with those obtained from analysis of similar models using system size expansions of the master equation.  In all of these analyses, the discrete nature of predator prey populations and finite size effects lead to persistent oscillations in time, but spatial patterns fail to form.  The path integral formalism goes beyond mean field theory and provides a focus on individual realizations of the stochastic time evolution of population not captured in the standard master equation approach.
\end{abstract}



\maketitle

When constructing models of biological phenomenon, observations of stable, periodic behavior have generally been taken to imply that the model will contain a stable limit cycle.  In the context of ecological modeling, both simple heuristic arguments and field observations support predator-prey oscillations in ecosystems.  However, the simple differential equation (mean field) models of predator prey dynamics do not exhibit limit cycles \cite{NOWA06, STROG94}.  Several authors have addressed this difficulty by developing spatial individual level models (ILMs) that incorporate the stochastic effects of individual predator-prey interactions as in, for example, \cite{MOBI06,MOBI07,BOCC94,ANTA01}.  These models yield limit cycles \cite{ANTA01} or stochastically induced cycles dependent on space \cite{BOCC94,MOBI06,MOBI07}.  However, recent work on a 0 dimensional model has shown that intrinsic noise without space is sufficient to generate temporal oscillations in predator-prey populations \cite{MCKAN05}.  Generalization of this work to space shows oscillations in time, but fails to exhibit oscillations in space \cite{LUGO08}.

The purpose of the present work is to develop a modified version of the spatial ILM of predator-prey interactions in \cite{LUGO08} and analyze the oscillatory fluctuations using path integral techniques.  Our model includes the motion of both predator and prey, does not have a hard constraint on the number of organisms that can be present in a patch and will be found to have oscillations at the global scale consistent with previous results \cite{LUGO08}.  We map the master equation to a bosonic field theory \cite{DOI76, MIKH81, PELI85, GOLD84} to obtain a simple derivation of coupled Langevin equations for the fluctuations of predator-prey populations. 

\section{Definition of the model and master equation}

Consider a single, well-mixed patch of volume $V$.  Species $A$ is a predator for species $B$.  We then have the following reactions:
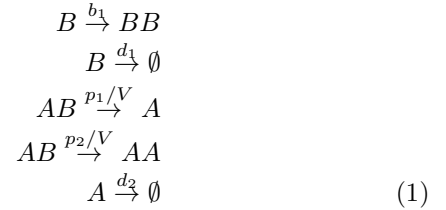
\begin{eqnarray}
B\stackrel{b_1}{\rightarrow}BB \nn \\
B\stackrel{d_1}{\rightarrow} \emptyset \nn \\
AB\stackrel{p_1/V}{\rightarrow}A \nn \\
AB\stackrel{p_2/V}{\rightarrow}AA \nn \\
A\stackrel{d_2}{\rightarrow} \emptyset
\label{1}
\end{eqnarray}

We give the rates of the two body reactions an inverse $V$ dependence, which is interpreted as the volume scaling of the probability in a volume $V$ that the two organisms will be close enough to interact. 

The above model contains a serious defect: in the absense of predation, the prey population diverges to infinity (in mean field).  Even with predators present, this defect manifests itself through the presence of non-generic, initial condition dependent oscillations. To overcome this defect, there exist a variety of options to induce a finite ``carrying capacity" for prey.  Each option has advantages depending on the predator-prey system being described, though many of the predictions end up being generic \cite{WASH07}.  One option is to restrict the total patch population to some number $N$, including empty space (i.e. $N_A+N_B+N_E=N$).  This is the ``urn model" description \cite{MCKAN04}.  In spatial models, $N$ is often chosen to be 1, which is equivalent to a coarse graining scheme which takes a patch to be the space required for one organism.  When $N>1$ models are generalized to space, a patch is a locally well mixed area.  Space is added as diffusion between such patches.  In our model, we adopt the perspective that a patch is a well mixed region with many organisms, but do not constrain the population to a given $N$, choosing instead to obtain a finite carrying capacity by allowing the death rate to increase with concentration. Equivalently, we could have simply included an intraspecies competition reaction.  An advantage of the current approach is that it avoids nonlinear diffusive cross terms in spatial urn models that do not seem to change the dynamics substantially from versions without the cross terms \cite{LUGO08}.  Additionally, urn models lead to complications in the interpretation of model parameters at the mean field level and in the master equation due to the fact that reaction rates in urn models must be combined with the joint probability for drawing the reactants from the urn prior to use in the master equation or mean field description leading to complex combinations of parameters \cite{MCKAN04}.  With the soft constraint applied here, the reaction rates have similar, predictable meanings at every level of description from master equation to mean field.

Formally, we include the concentration dependence of the death rate by noting that $n_A = N_A/V$ is small 

\begin{equation}
d_1(n_A) = d_1(0)+c n_A + O(n_A ^2),\:\:c=d'(0)>0 
\label{2}
\end{equation}

We can now write a master equation for the patch

\begin{eqnarray}
\partial_t P(m,n) = d_1(-nP(m,n) + (n+1)P(m,n+1))\nn \\ +c(-n^2P(m,n)+ (n+1)^2P(m,n+1)) \nn \\
+b_1(-nP(m,n)+(n-1)P(m,n-1))\nn \\ +p_1(-mnP(m,n)+(n+1)mP(m,n+1))+\nn \\ p_2(-mnP(m,n)+(m-1)(n+1)P(m-1,n+1))+\nn \\ d_2(-mP(m,n)+(m+1)P(m+1,n)) \:\:
\label{3}
\end{eqnarray}

Where $m$ denotes the number of predators, and $n$ denotes the number of prey.  This master equation defines the time evolution of the 
probability distribution of population states.

\section{Mapping to path integral formulation}

To analyze the predator prey dynamics, we map Eq. \ref{3} to a field theory.  This is done using the standard Doi formalism to obtain a second quantized Hamiltonian \cite{DOI76} and bosonic coherent states to map the resulting theory to a path integral.  For our approach and helpful reviews, see \cite{MATT98, CARD96}.  The mapping is achieved by introducing the state vector
\begin{equation}
| \psi \rangle = \sum_{m,n} P(m,n)|m,n\rangle
\end{equation}
and the operator pairs $a,\hat{a}$, $b,\hat{b}$ such that
\begin{eqnarray}   
a|m,n\rangle = m|m-1,n \rangle \nn \\
\hat{a}|m,n\rangle = |m+1,n\rangle \nn \\
\left[ a, \hat{a} \right] = 1	\nn \\
b|m,n\rangle = n|m,n-1 \rangle \nn \\
\hat{b}|m,n\rangle = |m,n+1\rangle \nn \\
\left[ b, \hat{b} \right] = 1 \:\:
\label{4}
\end{eqnarray}
Finally, all other commutators are zero.  We can then rewrite the dynamics given by the master equation (Eq. \ref{3}) as a Schrodinger like equation.
\begin{equation}
\partial_t |\psi \rangle = - \hat{H}(a,\hat{a},b,\hat{b})|\psi\rangle
\label{5}
\end{equation}

We now can now specify the Hamiltonian (more accurately Liouvillian \cite{GOLD84}) operator by multiplying the master equation by the state vector $|m,n\rangle$, summing over $m$ and $n$, and applying the algebra of Eq. \ref{5} to replace $m$ and $n$ by various combinations of the operators $a, \; \hat{a}$ and $b, \; \hat{b}$.  From this algebra, working out the structure of the Hamiltonian is direct and simple.  As an example, we work out the term corresponding to prey birth explicitly

\begin{eqnarray}
b_1\sum_{m,n}(-nP(m,n)+(n-1)P(m,n-1))|m,n\rangle \nn \\
= b_1\sum_{m,n}(-\hat{b}b P(m,n)+(n-1)P(m,n-1))|m,n\rangle \nn \\
= -b_1\hat{b}b|\psi \rangle + \sum_{m,n}nP(m,n)|m,n+1\rangle \nn \\
= -b_1\hat{b}b|\psi \rangle + b_1 \hat{b}\hat{b}b|\psi\rangle \:\:
\label{6}
\end{eqnarray}

Other terms are treated analogously.  With normal ordering, this leads to the Hamiltonian 

\begin{eqnarray}
\hat{H}= b_1(\hat{b}b-\hat{b}^2b)+d_1(\hat{b}b-b)+\frac{c}{V}(\hat{b}^2b^2-\hat{b}b^2) \nn
\\+\frac{p_1}{V}(\hat{a}a\hat{b}b-\hat{a}ab)+\frac{p_2}{V}(\hat{a}a\hat{b}b-\hat{a}^2ab) \nn \\
+d_2(\hat{a}a-a)
\label{7}
\end{eqnarray}

Expectation values of functions of the random variables $m$ and $n$ are given by

\begin{eqnarray}
\langle f \rangle = \langle 0,0 | e^{a+b} f(\hat{a},a,\hat{b},b)e^{-H(\hat{a},a,\hat{b},b)t} |\psi(0)\rangle \nn \\
\label{8}
\end{eqnarray}
  
Using bosonic coherent states, we write Eq. \ref{8} as a path integral resulting in a Lagrangian description of the dynamics with generalization to space \cite{MIKH81, PELI85}.  Since we are interested in persistent oscillations around the only stable fixed point in the system, our choice of initial conditions is irrelevant and can be ignored. To link patches together for a spatial description, we define a lattice of patches and demand that each organism carry out a random walk on the lattice with given hopping probabilities for predator and prey.  The continuum limit of a random walk is well known to be diffusion.  We thus define diffusion rates $D_1$ and $D_2$ for predator and prey respectively and add diffusion operators to the Lagrangian.  Careful manipulation of the field operators leads to the same results, provided the hopping probability for a species $\tau$ scales as $\tau \sim 1/a^2$ where $a$ is the lattice constant taken to 0 in the continuum limit.  Then $D=\lim_{a\to 0} a^2 \tau$.  The resulting Lagrangian density is given by

\begin{eqnarray}  
\mathcal{L} = a^*\partial_t a + b^* \partial_t b - D_1 a^* \nabla^2 a - D_2 b^*\nabla^2 b \nn \\
+ H (\hat{b}, \hat{a}, b, a)
\label{9}
\end{eqnarray}

With fields derived from boson operators, the Lagrangian form of the master equation is not simply interpreted.  This is because the field variables in the Lagrangian are not simply related to the physical variables of population number.  This proves to be the source of difficulties in deriving correlation functions that are physically meaningful.  To address this difficulty, we use a standard semi canonical Cole-Hopf transformation \cite{JANS08} to transform the field variables to density variables

\begin{eqnarray}
a=z e^{-\hat{z}}, \:
\hat{a} = e^{\hat{z}} \\
b=\rho e^{-\hat{\rho}}, \:
\hat{b} = e^{\hat{\rho}}
\label{10}
\end{eqnarray}

This formulation has the advantage that $z$ and $\rho$ can be directly interpreted as the density variables for predator and prey respectively, while $\hat{\rho}$ and $\hat{z}$ generate noise terms at quadratic order.  The transformed Lagrangian takes the form

\begin{eqnarray}
\mathcal{L}=\hat{z}\partial_t z +\hat{\rho}\partial_t \rho -D_1\hat{z}\nabla^2 z -D_1z(\nabla\hat{z})^2 \nn \\
-D_2 \rho (\nabla \hat{\rho})^2 - D_2 \hat{\rho}\nabla^2 \rho - b_1\rho(1-e^{\hat{\rho}}) \nn \\
+ d_1\rho(1-e^{-\hat{\rho}})+\frac{c}{V}\rho^2(1-e^{-\hat{\rho}}) \nn \\
+\frac{p_1}{V}z\rho(1-e^{-\hat{\rho}})+\frac{p_2}{V}z\rho(1-e^{\hat{z}-\hat{\rho}}) \nn \\
+d_2 z(1-e^{-\hat{z}})
\label{11}
\end{eqnarray}

In this form, the Lagrangian has diffusive noise, and difficult to handle exponential terms.  In the following section, we exploit the small parameter $1/V$ to resolve these difficulties and analyse the theory. 


\section {Derivation of Mean field theory and quasi-cycles from Large $V$ expansion}

From the Lagrangian in Eq. \ref{11}, we can proceed directly by rewriting the fields as

\begin{eqnarray}
\hat{z} \rightarrow \frac{\hat{z}}{\sqrt{V}}, \; \;
\hat{\rho} \rightarrow \frac{\hat{\rho}}{\sqrt{V}} \nn \\
z=V\varphi +\sqrt{V}\eta, \;\;
\rho = V \phi + \sqrt{V}\xi
\label{12}
\end{eqnarray}

\noindent and inserting them into the Lagrangian.  These forms are intended to capture Gaussian fluctuations in the spirit of the traditional system size expansion of the master equation \cite{VANK92} while directly manipulating the population variables.  The fields $\hat{z}$ and $\hat{\rho}$ have a mean field value of $0$ due to conservation of probability \cite{CARD96}.  This means that within the Gaussian approximation, the leading order term in those fields is a small correction of order $1/\sqrt{V}$ as above.  

To derive the mean field theory and the fluctuations, we then insert the rhs forms of the fields in Eq. \ref{12} into the Lagrangian Eq. \ref{11} and retain only leading and next to leading order, resulting in an effective Lagrangian of the form

\begin{equation}
\mathcal{L}=\sqrt{V}\mathcal{L}_1+\mathcal{L}_2 +O(1/\sqrt{V})
\label{13}
\end{equation}

Deriving each of these terms is straightforward.  For purposes of illustration, we will carry out the expansion for the prey birth term explicitly

\begin{eqnarray}
b_1\rho(1-e^{\hat{\rho}})  \nn \\
=b_1(V \phi + \sqrt{V}\xi)(-\frac{\hat{\rho}}{V}-\frac{\hat{\rho}^2}{2V}) \nn \\
=b_1(-\sqrt{V}\hat{\rho}\phi-\frac{\hat{\rho}^2\phi}{2}-\hat{\rho}\eta)
\label{14}
\end{eqnarray}

Carrying this out for each term in the Lagrangian and collecting terms yields at order $\sqrt{V}$

\begin{eqnarray}
\mathcal{L}_1 = \hat{\rho} \partial_t \phi +\hat{z} \partial_t \varphi -D_1 \hat{z} \nabla^2 \varphi -D_2 \hat{\rho} \nabla^2 \phi \nn \\
-b_1\phi \hat{\rho} +d_1\varphi \hat{\rho} +c \hat{\rho}\phi^2 +p_1\hat{\rho} \varphi \phi +p_2\hat{\rho} \phi \varphi \nn \\
-p_2\hat{z} \phi \varphi +d_2\hat{z} \varphi 
\label{15}
\end{eqnarray}

Minimizing this term provides the mean field theory.  For $V \rightarrow \infty$, this minimum is exact.  The Euler-Lagrange equations are:

\begin{eqnarray}
\frac{\delta \mathcal{L}_1}{\delta \hat{z}} = \partial_t \varphi -D_1 \nabla^2 \varphi -p_2 \phi \varphi +d_2 \varphi = 0 \nn \\
\frac{\delta \mathcal{L}_1}{\delta \hat{\rho}} = \partial_t \phi -D_2 \nabla^2 \phi -b_1\phi +d_1\phi +c \phi^2 \nn \\
+p_1 \varphi \phi +p_2 \phi \varphi = 0
\label{16}
\end{eqnarray}

These are the standard Lotka-Volterra equations generalized to include space.  They do not satisfy the criteria for pattern formation in predator-prey equations (reviewed in \cite{HOLM94}), which generically require more complex predation interactions.  The long time dynamics relax to spatially uniform predator-prey populations with magnitudes given by the fixed points of the ordinary differential equations obtained by dropping the diffusion operator in Eqs. \ref{16} above.  

At next to leading order, we fourier transform and switch to matrix notation, defining

\begin{eqnarray}
\bf{x} = \left( \begin{array}{c} \eta \\   
				 	\xi   \end{array} \right),\				 					     			
\bf{y}=	\left( \begin{array}{c} \hat{z} \\
				 				\hat{\rho}   \end{array} \right)			
\label{17}
\end{eqnarray}

By simply collecting terms as in Eq. \ref{12} we can write down $\mathcal{L}_2$ as
 
\begin{equation}
\mathcal{L}_2 = i\omega \bf{y}^T \bf{x} + \bf{y}^T\bf{A}\bf{x}-\frac{1}{2}\bf{y}^T\bf{B}\bf{y}
\label{19}
\end{equation}

The matrices are given by

\begin{eqnarray}
\bf{A} = \left( \begin{array}{cc} D_1 k^2 & -p_2 \varphi \\
																	(p_1+p_2)\phi				 & D_2 k^2 +c\phi 
																	\end{array} \right)
\label{20}
\end{eqnarray}

\noindent and 

\begin{eqnarray}
\bf{B} =  
\left( \begin{array}{cc}2(d_2 + D_1 k^2)\varphi & -p_2\varphi\phi \\
				 -p_2\varphi\phi & 2(b_1+D_2 k^2)\phi 
				 \end{array} \right) 
\label{21}
\end{eqnarray}
  
We now note that the vector $\bf{y}$ is a response field in the Martin Siggia Rose response function formalism for Langevin equations \cite{MART73, JANS76}.  Thus the fluctuations around mean field in the path integral are coupled Langevin equations.  The resulting Langevin equations with the appropriate noise and correlations are

\begin{eqnarray}
-i\omega \bf{x} = \bf{A}\bf{x} +\bf{\gamma(\omega)} \nn \\
\langle \gamma_i (\omega) \gamma_j(-\omega)\rangle = B_{ij}
\label{22}
\end{eqnarray}

These equations are of the same form as the equations reported in \cite{MCKAN05,MCKAN07} and are easily solved using simple linear algebra manipulations \cite{MCKAN07}  

\begin{eqnarray}
\bf{x} = -(\bf{A} + i \omega)^{-1}\bf{\gamma(\omega)} \equiv \bf{D(\omega)}^{-1}\bf{\gamma(\omega)} \nn \\
\rightarrow x_1= \eta = -det(\bf{D})^{-1}(D_{1 1}\gamma_1-D_{1 2}\gamma_2) \nn \\
x_2= \xi = -det(\bf{D})^{-1}(D_{2 1}\gamma_1-D_{2 2}\gamma_2)
\label{23}
\end{eqnarray}

To obtain information from these solutions, we calculate the average power spectrum which captures oscillations but is free of phase cancellations \cite{MCKAN05}.  The average power spectrum is obtained by taking the amplitude squared and averaging.  For predator fluctuations this gives

\begin{eqnarray}
\langle x_1 x_1^*\rangle = \frac{\alpha_k+\beta_k\omega^2}{(\omega^2-\Omega_k^2)^2+\Gamma_k^2\omega^2}\nn \\
\label{24}
\end{eqnarray}

\noindent with 

\begin{eqnarray}
\alpha_k=B_{11}(k) A_{22}^2+B_{22}(k)A_{12}^2 \nn \\
\beta_k=B_{11}(k) \nn \\
\Omega_k^2=D_1 k^2(D_2 k^2 +c\phi)+p_2(p_1+p_2)\phi \varphi>0 \nn \\
\Gamma=-A_{11}-A_{22}
\label{25}
\end{eqnarray}

The power spectrum contains a nontrivial peak in $\omega$ corresponding to the expected temporal oscillations.  The peak in $k$ is at 0 wavenumber as can be seen from the strictly increasing functions of $k$ present in the spectrum.  This rules out spatial pattern formation.  These results are in qualitative agreement with results from expansion of the master equation Urn models \cite{MCKAN05,LUGO08}.  Additional work will investigate the scaling of population fluctuations near extinction transitions and in disordered environments.  These applications are of clear ecological interest and are difficult to study with system size expansions.  However, they can be studied using well known methods from field theory in the functional integral formalism.

We thank Nigel Goldenfeld for suggesting this problem and for helpful discussions.  This work was partially supported by the National Science Foundation and the Department of Energy grants NSF-EF-0526747 and DOE-2005-05818.

\bibliographystyle{apsrev}

\bibliography{predatorprey}

\begin{thebibliography}{22}
\expandafter\ifx\csname natexlab\endcsname\relax\def\natexlab#1{#1}\fi
\expandafter\ifx\csname bibnamefont\endcsname\relax
  \def\bibnamefont#1{#1}\fi
\expandafter\ifx\csname bibfnamefont\endcsname\relax
  \def\bibfnamefont#1{#1}\fi
\expandafter\ifx\csname citenamefont\endcsname\relax
  \def\citenamefont#1{#1}\fi
\expandafter\ifx\csname url\endcsname\relax
  \def\url#1{\texttt{#1}}\fi
\expandafter\ifx\csname urlprefix\endcsname\relax\def\urlprefix{URL }\fi
\providecommand{\bibinfo}[2]{#2}
\providecommand{\eprint}[2][]{\url{#2}}

\bibitem[{\citenamefont{Nowak}(2006)}]{NOWA06}
\bibinfo{author}{\bibfnamefont{M.~A.} \bibnamefont{Nowak}},
  \emph{\bibinfo{title}{Evolutionary Dynamics}}
  (\bibinfo{publisher}{Belknap/Harvard Press}, \bibinfo{year}{2006}).

\bibitem[{\citenamefont{Strogatz}(1994)}]{STROG94}
\bibinfo{author}{\bibfnamefont{S.~H.} \bibnamefont{Strogatz}},
  \emph{\bibinfo{title}{Nonlinear Dynamics and Chaos}}
  (\bibinfo{publisher}{Westview Press}, \bibinfo{year}{1994}).

\bibitem[{\citenamefont{Mobilia et~al.}(2006)\citenamefont{Mobilia, Georgiev,
  and Tauber}}]{MOBI06}
\bibinfo{author}{\bibfnamefont{M.}~\bibnamefont{Mobilia}},
  \bibinfo{author}{\bibfnamefont{I.~T.} \bibnamefont{Georgiev}},
  \bibnamefont{and} \bibinfo{author}{\bibfnamefont{U.~C.}
  \bibnamefont{Tauber}}, \bibinfo{journal}{Phys.\ Rev.~E}
  \textbf{\bibinfo{volume}{73}}, \bibinfo{pages}{040903(R)}
  (\bibinfo{year}{2006}).

\bibitem[{\citenamefont{Mobilia et~al.}(2007)\citenamefont{Mobilia, Georgiev,
  and Tauber}}]{MOBI07}
\bibinfo{author}{\bibfnamefont{M.}~\bibnamefont{Mobilia}},
  \bibinfo{author}{\bibfnamefont{I.~T.} \bibnamefont{Georgiev}},
  \bibnamefont{and} \bibinfo{author}{\bibfnamefont{U.~C.}
  \bibnamefont{Tauber}}, \bibinfo{journal}{J.\ Stat.\ Phys.}
  \textbf{\bibinfo{volume}{128}}, \bibinfo{pages}{447} (\bibinfo{year}{2007}).

\bibitem[{\citenamefont{Boccara et~al.}(1994)\citenamefont{Boccara, Roblin, and
  Roger}}]{BOCC94}
\bibinfo{author}{\bibfnamefont{N.}~\bibnamefont{Boccara}},
  \bibinfo{author}{\bibfnamefont{O.}~\bibnamefont{Roblin}}, \bibnamefont{and}
  \bibinfo{author}{\bibfnamefont{M.}~\bibnamefont{Roger}},
  \bibinfo{journal}{Phys.\ Rev.~E} \textbf{\bibinfo{volume}{50}},
  \bibinfo{pages}{4531} (\bibinfo{year}{1994}).

\bibitem[{\citenamefont{Antal and Droz}(2001)}]{ANTA01}
\bibinfo{author}{\bibfnamefont{T.}~\bibnamefont{Antal}} \bibnamefont{and}
  \bibinfo{author}{\bibfnamefont{M.}~\bibnamefont{Droz}},
  \bibinfo{journal}{Phys.\ Rev.~E} \textbf{\bibinfo{volume}{63}},
  \bibinfo{pages}{056119} (\bibinfo{year}{2001}).

\bibitem[{\citenamefont{McKane and Newman}(2005)}]{MCKAN05}
\bibinfo{author}{\bibfnamefont{A.~J.} \bibnamefont{McKane}} \bibnamefont{and}
  \bibinfo{author}{\bibfnamefont{T.~J.} \bibnamefont{Newman}},
  \bibinfo{journal}{Phys.\ Rev.\ Lett.} \textbf{\bibinfo{volume}{94}},
  \bibinfo{pages}{218102} (\bibinfo{year}{2005}).

\bibitem[{\citenamefont{Lugo and McKane}(2008)}]{LUGO08}
\bibinfo{author}{\bibfnamefont{C.}~\bibnamefont{Lugo}} \bibnamefont{and}
  \bibinfo{author}{\bibfnamefont{A.~J.} \bibnamefont{McKane}},
  \bibinfo{journal}{arXiv:0806.1287v1 [q-bio.PE]}  (\bibinfo{year}{2008}).

\bibitem[{\citenamefont{Doi}(1976)}]{DOI76}
\bibinfo{author}{\bibfnamefont{M.}~\bibnamefont{Doi}}, \bibinfo{journal}{J.
  Phys. A.} \textbf{\bibinfo{volume}{9}}, \bibinfo{pages}{1465}
  (\bibinfo{year}{1976}).

\bibitem[{\citenamefont{Mikhailov}(1981)}]{MIKH81}
\bibinfo{author}{\bibfnamefont{A.~S.} \bibnamefont{Mikhailov}},
  \bibinfo{journal}{Phys. Lett.} \textbf{\bibinfo{volume}{85}},
  \bibinfo{pages}{214} (\bibinfo{year}{1981}).

\bibitem[{\citenamefont{Peliti}(1985)}]{PELI85}
\bibinfo{author}{\bibfnamefont{L.}~\bibnamefont{Peliti}}, \bibinfo{journal}{PJ.
  Physique} \textbf{\bibinfo{volume}{46}}, \bibinfo{pages}{1469}
  (\bibinfo{year}{1985}).

\bibitem[{\citenamefont{Goldenfeld}(1984)}]{GOLD84}
\bibinfo{author}{\bibfnamefont{N.}~\bibnamefont{Goldenfeld}},
  \bibinfo{journal}{J. Phys. A} \textbf{\bibinfo{volume}{17}},
  \bibinfo{pages}{2807} (\bibinfo{year}{1984}).

\bibitem[{\citenamefont{Washenberger et~al.}(2007)\citenamefont{Washenberger,
  Mobilia, and Tauber}}]{WASH07}
\bibinfo{author}{\bibfnamefont{M.~J.} \bibnamefont{Washenberger}},
  \bibinfo{author}{\bibfnamefont{M.}~\bibnamefont{Mobilia}}, \bibnamefont{and}
  \bibinfo{author}{\bibfnamefont{U.~C.} \bibnamefont{Tauber}},
  \bibinfo{journal}{J. Phys. Condens. Matter} \textbf{\bibinfo{volume}{19}},
  \bibinfo{pages}{065139} (\bibinfo{year}{2007}).

\bibitem[{\citenamefont{McKane and Newman}(2004)}]{MCKAN04}
\bibinfo{author}{\bibfnamefont{A.~J.} \bibnamefont{McKane}} \bibnamefont{and}
  \bibinfo{author}{\bibfnamefont{T.~J.} \bibnamefont{Newman}},
  \bibinfo{journal}{Phys.\ Rev.~E} \textbf{\bibinfo{volume}{70}},
  \bibinfo{pages}{041902} (\bibinfo{year}{2004}).

\bibitem[{\citenamefont{Mattis and Glasser}(1998)}]{MATT98}
\bibinfo{author}{\bibfnamefont{D.}~\bibnamefont{Mattis}} \bibnamefont{and}
  \bibinfo{author}{\bibfnamefont{M.~L.} \bibnamefont{Glasser}},
  \bibinfo{journal}{Rev.\ Mod.\ Phys.} \textbf{\bibinfo{volume}{70}},
  \bibinfo{pages}{979} (\bibinfo{year}{1998}).

\bibitem[{\citenamefont{Cardy}(1996)}]{CARD96}
\bibinfo{author}{\bibfnamefont{J.}~\bibnamefont{Cardy}},
  \bibinfo{journal}{arXiv:cond-mat/9607163v2}  (\bibinfo{year}{1996}).

\bibitem[{\citenamefont{Janssen and Tauber}(2005)}]{JANS08}
\bibinfo{author}{\bibfnamefont{H.~K.} \bibnamefont{Janssen}} \bibnamefont{and}
  \bibinfo{author}{\bibfnamefont{U.~C.} \bibnamefont{Tauber}},
  \bibinfo{journal}{Annals of Physics} \textbf{\bibinfo{volume}{315}},
  \bibinfo{pages}{147} (\bibinfo{year}{2005}).

\bibitem[{\citenamefont{Van~Kampen}(1992)}]{VANK92}
\bibinfo{author}{\bibfnamefont{N.~G.} \bibnamefont{Van~Kampen}},
  \emph{\bibinfo{title}{Stochastic Processes in Physics and Chemistry}}
  (\bibinfo{publisher}{Elsevier, New York}, \bibinfo{year}{1992}).

\bibitem[{\citenamefont{Holmes et~al.}(1994)\citenamefont{Holmes, Lewis, Banks,
  and Veit}}]{HOLM94}
\bibinfo{author}{\bibfnamefont{E.~E.} \bibnamefont{Holmes}},
  \bibinfo{author}{\bibfnamefont{M.~A.} \bibnamefont{Lewis}},
  \bibinfo{author}{\bibfnamefont{J.~E.} \bibnamefont{Banks}}, \bibnamefont{and}
  \bibinfo{author}{\bibfnamefont{R.~R.} \bibnamefont{Veit}},
  \bibinfo{journal}{Ecology} \textbf{\bibinfo{volume}{75}}, \bibinfo{pages}{17}
  (\bibinfo{year}{1994}).

\bibitem[{\citenamefont{Martin et~al.}(1973)\citenamefont{Martin, Siggia, and
  Rose}}]{MART73}
\bibinfo{author}{\bibfnamefont{P.~C.} \bibnamefont{Martin}},
  \bibinfo{author}{\bibfnamefont{E.~D.} \bibnamefont{Siggia}},
  \bibnamefont{and} \bibinfo{author}{\bibfnamefont{H.~A.} \bibnamefont{Rose}},
  \bibinfo{journal}{Phys.\ Rev.\ A} \textbf{\bibinfo{volume}{8}},
  \bibinfo{pages}{423} (\bibinfo{year}{1973}).

\bibitem[{\citenamefont{Bausch et~al.}(1976)\citenamefont{Bausch, Janssen, and
  Wagner}}]{JANS76}
\bibinfo{author}{\bibfnamefont{R.}~\bibnamefont{Bausch}},
  \bibinfo{author}{\bibfnamefont{H.~K.} \bibnamefont{Janssen}},
  \bibnamefont{and} \bibinfo{author}{\bibfnamefont{H.}~\bibnamefont{Wagner}},
  \bibinfo{journal}{Z. Phys. B.} \textbf{\bibinfo{volume}{24}},
  \bibinfo{pages}{113} (\bibinfo{year}{1976}).

\bibitem[{\citenamefont{McKane et~al.}(2007)\citenamefont{McKane, Nagy, Newman,
  and Stefanini}}]{MCKAN07}
\bibinfo{author}{\bibfnamefont{A.}~\bibnamefont{McKane}},
  \bibinfo{author}{\bibfnamefont{J.~D.} \bibnamefont{Nagy}},
  \bibinfo{author}{\bibfnamefont{T.~J.} \bibnamefont{Newman}},
  \bibnamefont{and} \bibinfo{author}{\bibfnamefont{M.~O.}
  \bibnamefont{Stefanini}}, \bibinfo{journal}{J.\ Stat.\ Phys.}
  \textbf{\bibinfo{volume}{128}}, \bibinfo{pages}{165} (\bibinfo{year}{2007}).

\end{thebibliography}

\end{document}